\documentclass{article}

\usepackage{amsmath}
\usepackage{epsfig}
\usepackage{epsf}
\usepackage{psfrag}
\usepackage{natbib}
\usepackage{amssymb}
\usepackage[latin1]{inputenc}
\usepackage{rotating}
\usepackage{amsbsy}
\usepackage{lscape}
\usepackage{color}
\usepackage{latexsym }
\usepackage{amsfonts }
\usepackage{amsthm }
\usepackage{amscd}
\usepackage{graphicx}
\usepackage{epstopdf}
\usepackage{hyperref}
\usepackage{amsmath}
\usepackage{multirow}
\usepackage{amsbsy}
\usepackage{bbm}

\newtheorem{theorem}{Theorem}

\theoremstyle{definition}

\newtheorem{remark}{Remark}

\newcommand{\be}{\begin{equation}}
\newcommand{\ee}{\end{equation}}
\newcommand{\ba}{\begin{eqnarray}}
\newcommand{\ea}{\end{eqnarray}}
\newcommand{\bas}{\begin{eqnarray*}}
\newcommand{\eas}{\end{eqnarray*}}

\newcommand{\blue}{\color{black}}
\newcommand{\e}{ { \mathbb{E}}}

\newcommand{\yy}{{\bf{y}}}

\newcommand{\argmax}{{\mbox{argmax}}}

\newcommand{\var}{ {\mathbb{V}\rm ar }}
\newcommand{\cov}{ {\mathbb{C}\rm ov}}

\newcommand{\btheta}{\mbox{\boldmath $\theta$}}

\newcommand{\p}{\mbox{\sf Pr}}

\newcommand{\bq}{\mbox{\bf q}}

\newcommand{\bB}{\mbox{\bf B}}

\newcommand{\bH}{\mbox{\bf H}}
\newcommand{\bW}{\mbox{\bf W}}
\newcommand{\bY}{\mbox{\bf Y}}

\newcommand{\bfe}{\mbox{\bf e}}
\newcommand{\ind}{\mathbbm{1}}
\newcommand{\iid}{{\sc iid}}
\newcommand{\cdf}{{\sc cdf}}
\newcommand{\amse}{\mbox{\sc amse}}

\begin{document}

\date{}
\title{Composite empirical likelihood for multisample clustered data}
 \author{Jiahua Chen $^{a, b}$, \ Pengfei Li $^{c}$, \ Yukun Liu $^{d}$\footnote{
Corresponding author.   E-mail:
ykliu@sfs.ecnu.edu.cn.}, \ and James V. Zidek $^{b}$\\
$^a$ Research Institute of Big Data, University of Yunnan, China;
$^b$Department of Statistics,  \\
   University of British Columbia,
       Vancouver, BC,  Canada;
$^c$ Department of Statistics \\
and Actuarial Science,
University of Waterloo, Waterloo,   ON Canada;
$^d$ Key \\ Laboratory of
Advanced Theory and Application in Statistics and
 Data Science - MOE,\\
School of Statistics, East China Normal University, Shanghai, China
}

\maketitle

\abstract{
In many applications, data cluster.  Failing to take the cluster structure into consideration generally
leads to underestimated variances of point estimators and inflated type I   errors   in hypothesis tests.
Many circumstance-dependent approaches have been developed to handle clustered data.
A working covariance matrix may be used in generalized estimating
equations. One may throw out the cluster structure  and use   only the cluster means,
or explicitly model the cluster structure. Our interest is the case where multiple samples of
clustered data are collected, and the population quantiles are particularly
important. We  develop a composite empirical likelihood
for clustered data under a  density ratio model.  This approach
avoids parametric assumptions on the population distributions or
the cluster structure. It efficiently utilizes the common features of
the multiple populations and the exchangeability of the cluster members.
We also develop a cluster-based bootstrap method to provide
valid variance estimation and  to control the type I errors.
We examine the performance of the proposed method through
simulation experiments and illustrate its usage via a real-world example.


%\vskip0.7cm\noindent\textbf{AMS 2000 Subject Classifications:}
%\textit{Primary} \textbf{62G15}, \textit{Secondary} \textbf{62G20}.

\vskip0.7cm\noindent\textbf{Short title: }\textit{Composite empirical likelihood}
 \thispagestyle{empty}

\vfil

\newpage
\section{Introduction}

In many applications, data cluster.
In longitudinal studies, repeated measurements are taken on the
same object {\blue over time} \citep{diggle2002analysis}.
In neurosciences, clusters form when each experiment yields data
from multiple synapses \citep{galbraith2010study}.
Clustered data occur in studies of eyes, ears, knees, teeth, and
coronary arteries as well as in other medical research \citep{rosner2006wilcoxon}.
In survey sampling, we take multiple observations from the same
block or poll stations of a city \citep{lohr2009sampling}.
In forestry, agriculture, and other {\blue industries}, multiple units
may be taken from the same plot, tree, {\blue  or production} shift
\citep{verrill2015simulation}.

Failing to take the cluster structure into consideration generally leads to
underestimated variances of the point estimators
and inflated type I errors in hypothesis tests
\citep{datta2005rank,verrill2015simulation}.
Many circumstance-dependent strategies have been
developed to avoid {\blue such} potential pitfalls.
Longitudinal studies are generally concerned with
identifying important factors influencing the outcome of various
treatments. Effective and valid data analysis can be achieved
through generalized estimating equations with a working covariance matrix
\citep{zeger1986longitudinal}.
When the inference focuses exclusively on the population means,
one may throw out the cluster structure {\blue and analyze the
cluster means \citep{galbraith2010study} instead.}
One may choose to explicitly model the cluster effect.
For instance, random effects models, parametric or nonparametric,
are useful \citep{matteson2014nonparametric}.
Last but not least, one may initially regard the clustered data as
if they are independent and then take the clusters into consideration
when evaluating the uncertainty \citep{chandler2007}.
In many cases, existing methods can be straightforwardly adapted to
handle clustered data \citep{rosner2003incorporation,rosner2006wilcoxon}.

This research is motivated by a specific application,
but the general problem is of equal interest.
Recently, the potentially damaging effect
on lumber of factors such as climate change, forest fires,
and plagues of insects has drawn increased attention.
These factors together with how the log is processed
and the product sizes all impact
{\blue the  strength and stiffness of the resulting products.}
There is an urgent need to examine and update the ongoing lumber-quality
monitoring procedures in the American Society for Testing and Materials (ASTM)
Standard D1990 \citep{astm2002standard}.
Accordingly,
\cite{verrill2015simulation} examine eight statistical tests
proposed by scientists from the United States Department of Agriculture
Forest Products Laboratory to determine if they perform acceptably
when applied to test data collected for monitoring purposes.
Their investigation reveals that when the data are clustered,
these tests fail to control the type I error.

In this application, multiple samples are available because
the data collection is performed annually. The samples
are clustered because the sample units are obtained in bundles.
The parameters of particular interest are quantiles, {\blue which}  are crucial to the reliability of building structures.
Clearly, these features are shared by many applications.
To handle multiple samples of clustered data,
we propose a composite empirical likelihood (CEL)
based on a density ratio model (DRM).
This approach avoids the parametric distributional assumption
and assumptions about the cluster structure.
It efficiently utilizes the common features of
the multiple populations and the exchangeability of the cluster members.
We also develop a cluster-based bootstrap method to provide
variance estimation, confidence intervals (CIs),
and effective monitoring tests.
The validity of the proposed method is rigorously established.
In developing our new method, we use many techniques from the literature.
The empirical likelihood (EL),
the DRM, and their combination can be found in
\cite{owen2001empirical},
\cite{anderson1979multivariate}, and \cite{qin1997goodness}, {\blue respectively}.
The two- and multi-sample problems based on independent data
can be found in  \cite{cao2006}, \cite{tsao2015}, and \cite{chen2013quantile}.
We also wish to cite \cite{datta2005rank, chandler2007,
jasa.2010.tm08545, cao2012, ozturk2016quantile}, and \cite{li2017semiparametric},
who work on various problems related to clustered data
 based  on  a rich variety of models and tools.

The paper is organized as follows.
Section 2 introduces multisample clustered data,
the nonparametric random effects model,
the DRM-based CEL, and the cluster-based bootstrap.
Some asymptotic results are given.
Section 3 uses simulation experiments to demonstrate
the effectiveness of the DRM-based CEL approach,
the bootstrap CI, and the    testing  methods.
The effects of overfitting and misspecifying the DRM are also investigated
and found to be negligible.
Section 4 applies the proposed method to a real-data example, and
Section 5 gives a summary and discussion.
 The proofs are  given in the Supplementary Material.

\section{Nonparametric random effect, density ratio model,
and composite empirical likelihood}

We consider the data analysis problem where $m+1$ independent
samples of clustered data are collected:
\[
\{ \yy^\tau_{k, j} = (y_{k, j, 1}, \ldots, y_{k, j, d}):
k=0, 1, \ldots, m; j=1, 2, \ldots, n_k\}.
\]
 In a forestry application, $k$ marks the year, $j$ the lots, and
$d$ the number of pieces from this lot.
In medical studies,  $k=0$ or $1$  identifies the
treatment and control.
Often, the joint distribution of the cluster
members is exchangeable
\citep{datta2005rank,rosner2006wilcoxon}.
Let $F_k$ be the joint distribution of the cluster members in population $k$.
Exchangeability means that, when $d=3$ for instance,
\[
F_k(y_1, y_2, y_3) = F_k(y_2, y_1, y_3) = F_k(y_3, y_1, y_2) = \cdots
\]
for any ordering of $y_1, y_2$, and $y_3$. This flexible exchangeable nonparametric
$F_k$ neatly models the random effect.
The exchangeability assumption implies
\[
G_k(y) = F_k(y, \infty, \infty) = F_k(\infty, y, \infty) = F_k( \infty, \infty, y).
\]
In addition, $G_k$ is also the distribution of any member
of a cluster from $F_k$.

For inference on the mean of $G_k$, one may
avoid dealing with the cluster structure by working
on $\bar{y}_{k,j} = d^{-1} \sum_{l=1}^d y_{k,j,l}$ since
$\e\{\bar{y}_{k,j}\} = \e\{y_{k,j,l}\}$.
Clearly, {\blue the quantiles} of the distribution of $\bar{y}_{k,j}$
differ from {\blue those} of $y_{k,j,l}$.
Hence, this approach is not applicable to the quantiles.
The Wilcoxon test is often used to compare two {\blue distributions.
However, it detects only departures from  $\p(X < Y)=0.5$. }
Its effectiveness for comparing other
aspects of multiple populations is limited even when it is
{\blue extended} to clustered data
\citep{datta2005rank,rosner2006wilcoxon}.

Clearly, when $F_k$ is exchangeable, the empirical
distribution formed by $\{y_{k, j, l}: l=1, \ldots, d;  j=1, \ldots, n_k\}$
is an unbiased estimator of $G_k$. To improve the
estimation precision, parametric approaches can be
used subject to the risk of model misspecification, which can
be serious for low and high quantiles of $G_k$.
A compromise between model robustness
and efficiency is a semiparametric model,
the DRM introduced by \cite{anderson1979multivariate}:
\ba
\label{DRM2}
\frac{dG_k(y)}{dG_0(y)}
=
 \exp\{\btheta_k^\tau \bq(y) \}
\ea
for some preselected basis function $\bq(y)$ of dimension $q$
and unknown parameter vectors $\btheta_k$, $k=1, \ldots, m$.
In this setting, $G_0$ is unspecified. If $G_0$ is
standard normal and $\bq(y) = (1, y, y^2)^\tau$, then $G_k$
can be any normal distribution. Hence, the normal distribution
family is part of the DRM when $\bq(y) = (1, y, y^2)^\tau$. Similarly, the
gamma distribution family is part of the DRM when $\bq(y) = (1, y, \log y)^\tau$.
Clearly, the DRM is very flexible with a choice of ``large'' basis function $\bq(y)$.
In addition, the DRM permits a convenient EL-based data analysis.

Following \cite{owen2001empirical}, the likelihood contribution of each
observed cluster vector $\yy_{k, j}$ is $dF_k(\yy_{k, j}) = \p_k( \bY = \yy_{k, j})$,
where the subscript
in $\p_k$ indicates that the computation is under $F_k$.
If the components of $\bY$ or those of $F_k$ were independent,
we would have
\[
\p_k( \bY = \yy_{k, j})
= \prod_{l=1}^d \p_k( \bY_j = y_{k, j, l})
= \prod_{l=1}^d dG_k(y_{k, j, l}).
\]
The EL function
under the ``incorrect'' independence assumption is hence given by
\be
\label{compositeEL}
L(G_0, \ldots, G_m)
= \prod_{k, j} \left \{ \prod_{l=1}^d dG_k(y_{k, j, l}) \right \}.
\ee
Note that $G_k$ and $F_k$ are mutually determined if
the cluster members are independent.
The product  in \eqref{compositeEL} and the summations below
with respect to $\{k, j\}$ are over their full {\blue ranges.}

When the cluster members are dependent,
$\prod_{l=1}^d dG_k(y_{k, j, l})$ is a product of marginal
probabilities {\blue and   does} not equal $\p_k( \bY = \yy_{k, j})$.
It remains informative about the likeliness of
the candidate distribution $G_k$ but possibly with some efficiency loss.
Following \cite{lindsay1988composite},
$L$ in \eqref{compositeEL} is a CEL.
A composite likelihood generally leads to
model robustness and a simplified numerical solution.
The use of the composite likelihood has received considerable attention;
we refer to \cite{varin2011overview}
for an overview of its recent development.

Following the generic recommendation in \cite{owen2001empirical}, we restrict
the form of $G_0$ to
\[
G_0(y) = \sum_{k,j,l} p_{k,j,l}  \ind( y_{k, j, l} \leq y),
\]
where $\ind(\cdot)$ denotes the indicator function.
Under the DRM assumption, we have
\[
G_r(y)
= \sum_{k,j,l} p_{k,j,l}  \exp\{\btheta_r^\tau \bq(y_{k,j,l})\} \ind( y_{k, j, l} \leq y), \quad
r=0,1,\ldots, m,
\]
where $\btheta_0 = 0$. Since the $G_r$'s are distribution functions, we have
\be
\label{drm.constraint}
\sum_{k,j,l} p_{k,j,l}  [\exp\{\btheta_r^\tau \bq(y_{k,j,l})\}-1] = 0,
\ee
for $r = 0, 1, \ldots, m$.
The maximum CEL estimators of the $G_k$'s
maximize $L(G_0, \ldots, G_m)$ under constraints \eqref{drm.constraint}.

The CEL is algebraically identical
to the EL of $G_0, \ldots, G_m$ when $\{y_{k, j, l}: j=1, \ldots, n_k, l=1, \ldots, d\}$
is an  independent and identically distributed (\iid) sample from $G_k$.
This allows direct use of algebraic results from \cite{chen2013quantile},
\cite{keziou2008empirical}, and \cite{qin1997goodness}.
Let $\btheta^\tau = (\btheta_0^\tau, \btheta_1^\tau, \ldots, \btheta_m^\tau)$ and
\[
\ell_n(\btheta)
=
- \sum_{k, j, l}
\log \big [\sum_{r=0}^m \rho_{r} \exp\{\btheta_r^{\tau} \bq(y_{k,j,l})\} \big ]
+ \sum_{k, j,l}  \btheta_k^{\tau} \bq(y_{k,j,l})
\]
with $\rho_{r} = n_r/n$ and $n=\sum_{r=0}^mn_r$.
The profile log CEL function
\[
\tilde \ell_n(\btheta)=\argmax_{G_0} \log\{L(G_0, \ldots, G_m)\}
\]
subject to constraints \eqref{drm.constraint} shares its maximum point and value
with $\ell_n(\btheta)$;
we hence work with the algebraically much simpler $\ell_n(\btheta)$
and regard it as the profile log CEL.

Let the maximum CEL estimator be
$\hat\btheta = \arg\max_{\btheta} \ell_n(\btheta)$.
Given $\hat\btheta$, we have
\[
\hat p_{k,j,l}
=
\frac{1}{nd}\frac{1}{\sum_{r=0}^m\rho_{r}\exp\{\hat\btheta_r^{\tau} \bq(y_{k,j,l})\}  }.
\]
Subsequently, the maximum CEL estimator of $G_{r}(y)$
is given by
\[
\hat G_r(y)
=
\sum_{k,j,l} \hat p_{k,j,l} \exp \{ \hat{\btheta}_r^\tau \bq(y_{k,j,l})\} \ind (y_{k,j, l} \leq y).
\]

\subsection{Asymptotic properties of CEL estimate \label{asymp.xi}}

We first state some general and nonrestrictive conditions.

\begin{enumerate}
\item[]{\bf C1.}
The total sample size $n = \sum_{k=0}^m n_k \to \infty$,
and
$\rho_k = n_k/n$ remains a constant (or within the $n^{-1}$ range).

\item[]{\bf C2.}
$F_{k}(\yy)$ is exchangeable, i.e., for any permutation $\phi(\yy)$ of $\yy$,
$
F_k\big(\phi(\yy)\big) =F_{k}(\yy).
$

\item[]{\bf C3.}
The marginal distributions $G_k$  satisfy the DRM (\ref{DRM2}) with true
parameter value $\btheta^*$ and
$\int h(y; \btheta) d G_0 < \infty$ in a neighborhood of $\btheta^*$,
{where \(
h(y; \btheta)
=
\sum_{k=0}^m  \rho_{k} \exp\{ \btheta_k^{\tau} \bq(y)\}.
\)}

\item[]{\bf C4.}
The components of  $\bq(y)$ are continuous and stochastically
linearly independent, and the first component is one.
\end{enumerate}

Remark: Stochastic linear independence  means that no linear combinations
can be 0 with probability 1 under $G_0$.
The variance is positive definite when the first
component of $\bq(\cdot)$ is not included.

We need some notation before stating the asymptotic results.
Let
\[
h_k(y; \btheta)
=
\rho_{k} \exp\{ \btheta_k^{\tau} \bq(y)\}/h(y; \btheta)
\]
with $h(y; \btheta)$ defined in {\bf C3}.
We use the shorthand $h_k(y)=h_k(y; \btheta^*)$
where $\btheta^*$ is the true value.
Let $\delta_{rs}=1$ when $r=s$ and 0 otherwise, and
$\bar G(y)=\sum_{k=0}^m \rho_k G_{k}(y)$.
We further define
$\bB_r(y)$ to be an $(mq)$-dimensional vector with its $s$th segment being
$$
\bB_{rs}(y)=\int \{\delta_{rs} h_r(x)-h_r(x) h_s(x)\}\bq(x)\ind(x\leq y) d{\bar G}(x),
$$
and $\bB_r=\bB_r(\infty)$.
Let $\bW$ be an $(mq) \times (mq)$ block matrix with each block
a $q \times q$ matrix, and the $(r,s)$th block {\blue being}
$$
\bW_{rs}=\int \bq(y)\bq^{\tau}(y)
\{ h_r (y) \delta_{rs} - h_r(y)h_s(y) \} d\bar G(y).
$$
Further, let $\bfe_r$ be an $m \times 1$ vector with the $r$th component being 1
and the remaining components 0, {\blue and
\(
\bH(y) =  \big(h_1(y), h_2(y), \cdots, h_m(y)\big)^{\tau}.
\) }
Finally, we define
\(
\gamma_{rs}(x; y) = h_r(x) \ind (x \leq y)
+
 \{\bB_r(y)\}^\tau \bW^{-1}   \{ \bfe_s - \bH(x) \}\otimes \bq(x),
\)
where $\otimes$ denotes the Kronecker product.

\begin{theorem}
\label{varcov2}
Assume conditions C1--C4.
Then for any $0\leq r,s\leq m$ and two real numbers
$x$ and $y$ in the support of $G_0(y)$,
$$
\sqrt{n}\Big(\hat G_{r} (x) - G_{r}(x), \hat G_{s} (y) - G_{s}(y)\Big)^\tau
$$
is asymptotically  jointly bivariate normal with mean 0 and
variance-covariance matrix
\be
\left(
\begin{array}{cc}
\omega_{rr}(x, x) & \omega_{rs}(x, y) \\
\omega_{rs}(x,y) & \omega_{ss}(y, y)
\end{array}
\right ),
\ee
where
\bas
\omega_{rs}(x,y)
=
\frac{1}{d\rho_r\rho_s}    \sum_{k=0}^m \rho_k
\blue
\cov\Big(   \gamma_{rk}(y_{k,1,1}; x) ,
\gamma_{sk}(y_{k,1,1}; y) + (d-1)\gamma_{sk}(y_{k,1,2}; y) \Big).
\eas
\end{theorem}

\vspace{1ex}
Although the {\blue cluster} structure is not explicitly accommodated
in our approach, its effect is in $\omega_{rs}(x,y)$.
When $r=s$ and $x=y$, $\omega_{rr}(x,x)$ is the asymptotic variance
of $\hat G_r(x)$.
%{\blue
%\bas
%\var\{ \hat G_r (x)\}
%=
%\frac{1}{d \rho_r^2}
%\sum_{k=0}^m \rho_k \big \{
%\var \big (  \gamma_{rk}(y_{k,1,1}; x) \big )
%+
%(d-1) \cov \big (\gamma_{rk}(y_{k,1,1}; x), \gamma_{rk}(y_{k,1,2}; x\big )
%\big \}.
%\eas
%}
The extra term $ \cov \big (\gamma_{rk}(y_{k,1,1}; x), \gamma_{rk}(y_{k,1,2}; x) \big )$
in $\omega_{rr}(x,x)$
is generally positive, leading to a larger variance for the clustered data.

After the $G_r(y)$'s are properly estimated
and their joint limiting distribution obtained,
other population parameters such as the means,
variances, and quantiles of $G_r(y)$ can be estimated
accordingly.
However, their joint distributions are not always simple to obtain,
particularly for quantiles in the presence of clustered data.
Since the quantiles are of particular interest in many applications,
some additional effort is needed here.

For any $\alpha \in (0, 1)$, the $\alpha$-quantile of $G_r(y)$ is defined
to be
\[
G^{-1}_r(\alpha) = \inf_y \{ G_r(y) \geq \alpha \}.
\]
A natural estimator in the current context is hence
\(
\hat \xi_{r}= \hat G_r^{-1}(\alpha),
\)
and we call it the CEL quantile.
We use $\hat \xi_r=  \hat G_r^{-1}(\alpha_1)$ and
$\hat \xi_s=  \hat G_s^{-1}(\alpha_2)$ to represent two arbitrary CEL quantiles.

\begin{theorem}
\label{varxi2}
Assume conditions C1--C4 and that
\begin{enumerate}
\item[]{\bf C5.}
The density functions $g_r(y)$ of $G_r(y)$ are continuously differentiable and
positive in a neighborhood of the quantiles under consideration.
\end{enumerate}
Then
\(
\sqrt{n}(\hat \xi_{r} - \xi_r, \hat \xi_s - \xi_s)^\tau
\)
is jointly asymptotically bivariate normal with mean 0 and variance-covariance matrix
\be
\Sigma_{rs}=
%\left(
%\begin{array}{cc}
%\sigma_{rr}& \sigma_{rs} \\
%\sigma_{rs} & \sigma_{ss}
%\end{array}
%\right )
%=
\left(
\begin{array}{cc}
\omega_{rr}(\xi_r, \xi_r)/g_r^2(\xi_r)& \omega_{rs}(\xi_r, \xi_s)/\{g_r(\xi_r)g_s(\xi_s)\} \\
 \omega_{rs}(\xi_r, \xi_s)/\{g_r(\xi_r)g_s(\xi_s)\}& \omega_{ss}(\xi_s, \xi_s)/g_s^2(\xi_s)
\end{array}
\right ).
\ee
\end{theorem}

\vspace{1ex}
Once $\omega_{rs}$ and $g_r(\cdot)$ are properly estimated,
asymptotically valid CIs and tests are conceptually
simple byproducts.
This approach, however, involves a delicate task of
searching for a suitable consistent and stable estimate of $\omega_{rs}$.
We avoid this task with a cluster-based bootstrap procedure \citep{efron1992bootstrap},
which is a general recipe for {\blue interval estimation and hypothesis testing.}

\subsection{Cluster-based bootstrapping method \label{bootstrap.xi}}

We propose a bootstrap procedure as follows.
Take a nonparametric random sample of $n_k$ clusters
from the $k$th sample {\blue $\{\yy_{k,j}^*, j = 1, \ldots, n_k\}$
for each $k =0, 1, \ldots, m$.}
Compute the maximum CEL estimator $\hat \btheta^*$
based on the bootstrapped sample.
Obtain the bootstrap CEL cumulative distribution function ({\cdf}) as
$
\hat G_r^*(y)
$ and the bootstrap version of the quantile estimator
$
\hat \xi_r^*=\inf\{y: \hat G_r^*(y)\geq\alpha\}
$.

For any functional of {\blue $G_r$ and  $G_s$}, $\varphi(G_r, G_s)$,
we compute its corresponding bootstrap value $\varphi(\hat G_r^*, \hat G_s^*)$.
Its conditional distribution, given data, can be simulated by the above
bootstrapping procedure.
This leads to a two-sided
$1-\gamma$ bootstrap interval estimate of  $\varphi(G_r, G_s)$:
$$
\mbox{BC}_p(\gamma)
=
\left[\tau_{n,\gamma/2}^*, \tau_{n,1-\gamma/2}^* \right]
$$
with $\tau^*_{n, \gamma}$ being the $\gamma$th bootstrap quantile
of the conditional distribution of $\varphi(\hat G_r^*, \hat G_s^*)$.
To test the hypothesis
\[
H_0: ~\varphi(G_r, G_s)= 0
\]
with size $\gamma$, we reject $H_0$ when the interval estimate
does not include $0$ in favor of the two-sided alternative hypothesis
$\varphi(G_r, G_s) \neq 0$, or when $\tau^*_{n, \gamma} > 0$ in favor
of the one-sided alternative hypothesis $\varphi(G_r, G_s) >  0$.

The following theorem validates the proposed bootstrap procedure.
The proof is tedious, and we cite Theorem 3.6 of
\cite{shao1995introduction} for a similar conclusion.

\begin{theorem}
\label{boot1}
Assume the conditions of Theorem 4 in the supplementary material
and some smoothness conditions on $\varphi(G_r, G_s)$.
Then, as $n \to \infty$,
\bas
&&\sup_{x}\Big|
{\p}^*
\left(
\sqrt{n}\{\varphi(\hat G_r^*, \hat G_s^*) - \varphi(\hat G_r, \hat G_s)\}\leq x
\right) \\
&& \hspace{3cm}
-
\p \left(\sqrt{n}\{\varphi(\hat G_r, \hat G_s) - \varphi(G_r, G_s)\}\leq x\right)
\Big |=
o_p(1)
\eas
where ${\p}^*$ denotes the conditional probability given data.
\end{theorem}

The result is presented as if $\varphi(\cdot)$ can
only be a function of two population distributions.
In fact, the general conclusion for multiple populations
is true, but the presentation can be tedious, and we therefore omit it.
In applications, the bootstrap percentiles $\tau^*$ are obtained via bootstrap
simulation. In the simulation study,
we used $B=9,999$ bootstrap samples to obtain the simulated $\tau^*$ values.

\section{Simulation}

We simulate data from two random effects models,
each consisting of four populations.
They represent two types of marginal distributions
with varying degrees of within-cluster dependence.

\vspace{1ex}
\noindent
{\it \bf Model 1: Normal random effects model.}
This model is also used by  \cite{verrill2015simulation}. {\blue
Let $y_{k,i,j}$ represent the strength of the $j$th piece of lumber in the $i$th cluster
from population $k$. We assume that
\begin{equation}
\label{random.effect}
y_{k,j,l} = \mu_k + \gamma_{k,j} + \epsilon_{k,j,l}
\end{equation}
for $l=1, 2, \ldots, d$,
where $\mu_k$ is the mean population strength,
$\gamma_{k,j}$ is the random effect of the $j$th cluster,
and $\epsilon_{k,j,l}$ is the error term.
The random effects and error terms
are normally distributed and independent of each other.
Because of the presence of $\gamma_{k,j}$,
the lumber strengths $y_{k,j,1}, y_{k,j,2}, \ldots$, $y_{k,j,d}$}
are correlated.
The populations in the model satisfy the DRM assumptions
with $\bq(y) = (1, y, y^2)^\tau$.

In the simulation, we generate data {\blue from} this model
with various choices of the parameter values.
One parameter setting is chosen to be $m+1 = 4$ with the population means
\[
\mu_0 = \mu_1 = 15.5, ~ \mu_2 = 14.7, ~ \mu_3 = 14.0;
\]
the variances of the random effect
\[
\sigma^2_{\gamma, 0} = \sigma^2_{\gamma, 1} = 1.44, ~
\sigma^2_{\gamma, 2} = 1.00, ~ \sigma^2_{\gamma, 3} = 1.00;
\]
and the error variance $\sigma^2_\epsilon = 4$.
The other parameter settings are in Table \ref{compare.mse1}.

\vspace{1ex}
\noindent
{\it \bf Model 2: Gamma random effects model.}
We use the multivariate gamma distributions defined in \cite{nadarajah2006some}
to create the next simulation model.
Let $U_1,\ldots,U_d$ be $d$ \iid\  random variables with beta distributions
having shape parameters $a$ and $b$ (positive constants) yielding a density
$$
\frac{\Gamma(a+b)}{\Gamma(a)\Gamma(b)}u^{a-1}(1-u)^{b-1} \ind(0 < u < 1).
$$
Further, let $W$ be a gamma-distributed random variable with
shape parameter $a+b$ and rate parameter $\beta$. Its distribution
has density function
$$
\frac{\beta^{a+b}w^{a+b-1}\exp(-\beta w)}{\Gamma(a+b)} \ind ( 0 < w).
$$
Let $\bY^\tau  =W\times (U_1, \ldots, U_d)$.
The distribution of $\bY$ is then the multivariate gamma
$MG(a, b, \beta)$ with correlation $Corr(Y_i,Y_j)=a/(a+b)$ for all $1 \leq i < j \leq d$.
The marginal distribution of $Y_1 = U_1W$ is gamma with shape parameter $a$ and
rate parameter $\beta$.
When $b=\infty$,  $Y_1,\ldots,Y_d$ become independent.
Populations under this model satisfy the DRM assumption
with  $\bq(y) = (1, y, \log y)^\tau$.

We choose $m+1 = 4$ populations with the parameter values
\bas
a_0&=& a_1 = 8, ~a_2 = 7, ~a_3 = 6; ~~
b_0=b_1=b_2=b_3=b;\\
\beta_0 &=& \beta_1 = 1, ~\beta_2 = 1.05, ~\beta_3 = 1.1.
\eas
In the simulation, clustered observations
are generated according to the multivariate gamma distribution
with the above parameters $a_k, \beta_k$. The value of $b$ will be
given later.

Under both models, the parameter values are chosen so that the means and
quantiles are equal in the first two populations
and lower in the third and fourth populations.
This allows us to determine the
type I errors based on the first two populations and
to compute the powers when comparing the first and third or fourth populations.
The population means and other characteristics are in good agreement
with the populations employed by  \cite{verrill2015simulation} and the real data sets.

\subsection{Composite EL and empirical quantiles }

Many aspects of population distributions are of potential interest,
but the most challenging tasks are estimation and {\blue testing} on population quantiles.
Hence, we focus on the effectiveness of the CEL quantiles.
Other parameters such as the population mean and variance
are much simpler to handle. The success on quantiles is most persuasive.

We use 10,000 repetitions to obtain
the average mean square errors (\amse s)
of the CEL quantiles and the
straight empirical (EMP) quantiles or their differences across the
four populations.
The simulation results for data generated from the two models
are presented in Tables \ref{compare.mse1} and \ref{compare.mse2}.
We simulated with $d=5$, $d=10$ and
various combinations of sample sizes, population
variances, and correlations.

\begin{table}[h]
\centering\caption{
\amse\ ($\times$100) of the composite EL and empirical quantiles
(CEL and EMP)  under normal random effects model. Here $\Delta \xi_{0,k,\alpha} = \xi_{0,\alpha} - \xi_{k,\alpha}$
\label{compare.mse1}}

\begin{center}
\tabcolsep 8pt
\renewcommand{\arraystretch}{1}

\begin{tabular}{c|rrrr rrrr}
\hline
&\multicolumn{4}{c }{$d=5$}&\multicolumn{4}{c}{$d=10$}\\
\hline
Method&\multicolumn{2}{|c}{\mbox{CEL}}&\multicolumn{2}{c }{\mbox{EMP}}
&\multicolumn{2}{c}{\mbox{CEL}}&\multicolumn{2}{c}{\mbox{EMP}}\\
$\alpha$ &0.05&0.10 &0.05&0.10  &0.05&0.10&0.05&0.10\\
\hline
&\multicolumn{8}{c}{\footnotesize $(n_0,n_1,n_2,n_3)=(25,30,40,40),
~~(\sigma^2_{\gamma,0},\sigma^2_{\gamma,1},\sigma^2_{\gamma,2},\sigma^2_{\gamma,3})
=(1.44,1.44,1.00,1.00)$} \\
$\xi_{0,\alpha}$  &18.31&14.64&25.58&18.65&12.84&10.60&16.77&12.72\\
$\xi_{2,\alpha}$  &10.01&7.72&14.08&9.78&6.53&5.17&8.41&6.16\\
$\xi_{3,\alpha}$  &10.90&8.11&13.79&9.74&6.97&5.40&8.51&6.17\\
$\Delta \xi_{0,1,\alpha}$&31.44&25.52&45.93&34.11&22.82&19.20&31.33&23.72\\
$\Delta \xi_{0,2,\alpha}$&27.21&22.05&40.54&28.81&18.35&15.23&25.21&18.67\\
$\Delta \xi_{0,3,\alpha}$&28.83&22.64&40.26&28.58&19.48&15.90&25.29&18.76 \\
\hline
&\multicolumn{8}{c}{\footnotesize$(n_0,n_1,n_2,n_3)=(25,30,40,40),
~~(\sigma^2_{\gamma,0},\sigma^2_{\gamma,1},\sigma^2_{\gamma,2},\sigma^2_{\gamma,3})
=(0.36,0.36,0.25,0.25)$}\\
$\xi_{0,\alpha}$  &10.37&8.04&16.13&11.39&6.08&4.82&9.05&6.49\\
$\xi_{2,\alpha}$  &6.90&5.07&10.52&6.68&3.78&2.87&5.37&3.67\\
$\xi_{3,\alpha}$  &7.74&5.50&10.30&6.88&4.28&3.10&5.51&3.75\\
$\Delta \xi_{0,1,\alpha}$&17.75&14.21&29.99&21.09&10.28&8.33&16.93&11.61\\
$\Delta \xi_{0,2,\alpha}$&16.04&12.59&26.53&17.96&9.57&7.65&14.51&10.31\\
$\Delta \xi_{0,3,\alpha}$&17.93&13.54&26.92&18.31&10.48&7.97&14.75&10.22\\
\hline
&\multicolumn{8}{c}{\footnotesize$(n_0,n_1,n_2,n_3)=(38,45,60,60),
~~(\sigma^2_{\gamma,0},\sigma^2_{\gamma,1},\sigma^2_{\gamma,2},\sigma^2_{\gamma,3})
=(1.44,1.44,1.00,1.00)$} \\
$\xi_{0,\alpha}$  &11.89&9.56&16.94&12.07&8.64&7.12&11.32&8.39\\
$\xi_{2,\alpha}$  &6.78&5.29&9.48&6.60&4.47&3.61&5.78&4.31\\
$\xi_{3,\alpha}$  &7.42&5.49&9.44&6.57&4.68&3.65&5.67&4.17\\
$\Delta \xi_{0,1,\alpha}$&20.21&16.59&30.66&21.91&14.92&12.64&20.71&15.46\\
$\Delta \xi_{0,2,\alpha}$&17.53&14.40&26.47&18.74&12.54&10.48&17.03&12.65\\
$\Delta \xi_{0,3,\alpha}$&19.03&15.13&26.07&18.77&13.41&10.83&17.23&12.59\\
\hline
&\multicolumn{8}{c}{\footnotesize$(n_0,n_1,n_2,n_3)=(38,45,60,60),
~~(\sigma^2_{\gamma,0},\sigma^2_{\gamma,1},\sigma^2_{\gamma,2},\sigma^2_{\gamma,3})
=(0.36,0.36,0.25,0.25)$}\\
$\xi_{0,\alpha}$  &6.85&5.30&11.05&7.27&4.01&3.17&6.01&4.26\\
$\xi_{2,\alpha}$  &4.61&3.43&6.78&4.53&2.52&1.93&3.55&2.46\\
$\xi_{3,\alpha}$  &5.22&3.69&6.87&4.59&2.85&2.03&3.63&2.45\\
$\Delta \xi_{0,1,\alpha}$&11.58&9.18&20.25&13.54&6.83&5.60&10.97&7.91\\
$\Delta \xi_{0,2,\alpha}$&10.80&8.50&17.84&11.85&6.18&4.95&9.64&6.68\\
$\Delta \xi_{0,3,\alpha}$&12.23&9.02&18.36&12.08&6.96&5.26&9.91&6.67\\
\hline
\end{tabular}

\end{center}
\end{table}

\begin{table}[h]
\centering\caption{\amse\ ($\times$100)  of the composite EL and empirical
quantiles (CEL and EMP) under gamma random effects model.
Here $\Delta \xi_{0,k,\alpha} = \xi_{0,\alpha} - \xi_{k,\alpha}$
\label{compare.mse2}}

\begin{center}
\tabcolsep 8pt
\renewcommand{\arraystretch}{1}

\begin{tabular}{c|rrrr rrrr}
\hline
&\multicolumn{4}{c }{$d=5$}&\multicolumn{4}{c}{$d=10$}\\
\hline
Method&\multicolumn{2}{c}{\mbox{CEL}}&\multicolumn{2}{c }{\mbox{EMP}}&\multicolumn{2}{c}{\mbox{CEL}}&\multicolumn{2}{c}{\mbox{EMP}}\\
$\alpha$ &0.05&0.10 &0.05&0.10  &0.05&0.10&0.05&0.10 \\
\hline
&\multicolumn{8}{c}{$(n_0,n_1,n_2,n_3)=(25,30,40,40),~~b=14$}\\
$\xi_{0,\alpha}$  &11.15&11.07&15.60&13.82&8.20&8.48&10.32&9.67\\
$\xi_{2,\alpha}$  &5.25&5.22&6.82&6.19&3.55&3.71&4.20&4.20\\
$\xi_{3,\alpha}$  &4.10&3.87&4.59&4.26&2.64&2.73&2.88&2.92\\
$\Delta \xi_{0,1,\alpha}$&19.42&19.59&28.15&25.63&14.56&15.15&18.64&17.87\\
$\Delta \xi_{0,2,\alpha}$&15.71&15.95&22.53&20.03&11.59&12.10&14.58&13.87\\
$\Delta \xi_{0,3,\alpha}$&15.19&14.92&20.28&17.93&10.90&11.19&13.34&12.52 \\
\hline
&\multicolumn{8}{c}{$(n_0,n_1,n_2,n_3)=(25,30,40,40),~~b=63$} \\
$\xi_{0,\alpha}$  &7.60&7.28&11.81&9.99&4.32&4.34&6.52&5.75\\
$\xi_{2,\alpha}$  &3.79&3.63&5.43&4.66&2.12&2.07&2.89&2.58\\
$\xi_{3,\alpha}$  &3.20&2.80&3.73&3.28&1.73&1.60&1.97&1.83\\
$\Delta \xi_{0,1,\alpha}$&12.75&12.55&21.52&18.39&7.26&7.46&11.87&10.39\\
$\Delta \xi_{0,2,\alpha}$&10.72&10.62&17.16&14.49&6.30&6.41&9.54&8.29\\
$\Delta \xi_{0,3,\alpha}$&10.94&10.28&15.85&13.35&6.15&6.01&8.61&7.55\\
\hline
&\multicolumn{8}{c}{$(n_0,n_1,n_2,n_3)=(38,45,60,60),~~b=14$}\\
$\xi_{0,\alpha}$  &7.40&7.36&10.02&9.27&5.35&5.55&6.73&6.54\\
$\xi_{2,\alpha}$  &3.46&3.47&4.43&4.15&2.44&2.54&2.91&2.89\\
$\xi_{3,\alpha}$  &2.72&2.58&3.07&2.83&1.73&1.75&1.91&1.90\\
$\Delta \xi_{0,1,\alpha}$&12.64&12.87&18.21&16.59&9.71&10.20&12.16&12.14\\
$\Delta \xi_{0,2,\alpha}$&10.75&10.85&14.72&13.54&7.74&8.09&9.68&9.42\\
$\Delta \xi_{0,3,\alpha}$&9.96&9.85&12.98&11.94&7.27&7.52&8.80&8.62 \\
\hline
&\multicolumn{8}{c}{$(n_0,n_1,n_2,n_3)=(38,45,60,60),~~b=63$}\\
$\xi_{0,\alpha}$             &5.01&4.81&7.83&6.61&2.81&2.80&4.28&3.71\\
$\xi_{2,\alpha}$              &2.53&2.40&3.66&3.12&1.41&1.40&1.94&1.76\\
$\xi_{3,\alpha}$              &2.10&1.85&2.52&2.16&1.14&1.07&1.33&1.20\\
$\Delta \xi_{0,1,\alpha}$&8.42&8.35&14.57&12.11&4.84&4.96&7.79&6.92\\
$\Delta \xi_{0,2,\alpha}$&7.25&7.10&11.47&9.80&4.09&4.16&6.25&5.50\\
$\Delta \xi_{0,3,\alpha}$&7.12&6.70&10.35&8.83&4.00&3.90&5.59&4.90\\
\hline
\end{tabular}

\end{center}
\end{table}

As expected, in all cases the CEL quantiles
have much lower \amse s than the corresponding
sample quantiles. The effectiveness of the CEL is evident.
We tried several versions of multivariate EL, and none of them
were as {\blue efficient} as the CEL presented here.
Because these results need lengthy preambles and add
little value, we choose to make only a simple remark here.

\subsection{Confidence intervals}

We next simulate the coverage precision of the CIs
constructed by the cluster-based bootstrap.
The approach is applicable generally, but we focus
on quantiles or quantile differences because of their importance
and because they stand for a sticky scenario.
CIs can also be obtained by the Wald method
in the form
$\hat \theta \pm z_{1-\alpha/2} \{\widehat {\var}(\hat \theta)\}^{1/2}$.
Bootstrap CIs are well known for giving better precision
in the coverage probabilities compared with Wald-type intervals
\citep{hall1988theoretical}, particularly when the normal approximation is poor.
Because of this, we do not attempt to show the superiority of the bootstrap interval.
Instead, {\blue we apply the Wald intervals to the empirical quantiles and
use  the  asymptotic  variance
\[
\widehat {\var} (\tilde \xi_r) = \frac{\alpha ( 1- \alpha)}{n_r d \hat g^2_r(\tilde \xi_r)},
\]
which is suitable only under an independence assumption,
and the corresponding $\widehat {\var} (\tilde \xi_r -\tilde \xi_s)$
in the Wald intervals for quantile differences.}
Here $\tilde \xi_r$ is the empirical quantile estimator of $\xi_r$
and $\hat g_r(y)$ is the kernel density estimator of $g_r(y)$,
in which the normal kernel is used and the bandwidth is selected by rule of thumb.
The anticipated poor performance of the Wald intervals illustrates
the danger of ignoring the cluster structure.

We generated data from the same models and used the
same parameter settings as in the previous section.
The simulated coverage probabilities
are summarized in Tables \ref{cover.pr.norm} and
\ref{cover.pr.gam} based on 10,000 repetitions.
The nominal level is 95\% and the simulation error is below $0.5\%$.

\begin{table}[h]
\centering\caption{Coverage probabilities (\%) of two-sided 95\% CIs under normal random effects model.
Here bootstrap composite EL and Wald empirical quantile intervals: CEL and EMP
\label{cover.pr.norm}}

\begin{center}
\tabcolsep 12pt
\renewcommand{\arraystretch}{1}

\begin{tabular}{c|rrrr rrrr}
\hline
&\multicolumn{4}{c }{$d=5$} &\multicolumn{4}{c}{$d=10$}\\
\hline
Method&\multicolumn{2}{c}{\mbox{CEL}}&\multicolumn{2}{c }{\mbox{EMP}}&\multicolumn{2}{c}{\mbox{CEL}}&\multicolumn{2}{c}{\mbox{EMP}}\\
$\alpha$ &0.05&0.10 &0.05&0.10  &0.05&0.10&0.05&0.10 \\
\hline
&\multicolumn{8}{c}{\footnotesize $(n_0,n_1,n_2,n_3)=(25,30,40,40),
~~(\sigma^2_{\gamma,0},\sigma^2_{\gamma,1},\sigma^2_{\gamma,2},\sigma^2_{\gamma,3})
=(1.44,1.44,1.00,1.00)$}\\
$\xi_{0,\alpha}$  &90.6&91.5&83.0&86.5&91.1&91.8&81.3&81.6\\
$\xi_{2,\alpha}$  &92.7&93.1&89.2&89.8&92.6&92.8&85.7&86.1\\
$\xi_{3,\alpha}$  &92.3&93.0&89.7&90.0&92.7&93.2&86.3&85.7\\
$\Delta \xi_{0,1,\alpha}$&94.2&94.1&86.8&88.1&93.7&93.6&82.5&81.9\\
$\Delta \xi_{0,2,\alpha}$&94.4&94.3&86.3&88.3&94.1&94.0&84.1&83.6\\
$\Delta \xi_{0,3,\alpha}$&94.5&94.2&86.5&88.5&94.3&94.0&83.9&83.4\\
\hline
&\multicolumn{8}{c}{\footnotesize $(n_0,n_1,n_2,n_3)=(25,30,40,40),
~~(\sigma^2_{\gamma,0},\sigma^2_{\gamma,1},\sigma^2_{\gamma,2},\sigma^2_{\gamma,3})
=(0.36,0.36,0.25,0.25)$}\\
$\xi_{0,\alpha}$  &92.0&92.5&88.0&91.2&92.6&92.8&89.0&90.7\\
$\xi_{2,\alpha}$  &93.1&93.6&92.0&93.2&93.4&93.8&91.4&91.6\\
$\xi_{3,\alpha}$  &92.9&93.3&91.9&92.9&93.4&93.8&91.2&91.9\\
$\Delta \xi_{0,1,\alpha}$&94.2&94.1&91.3&92.5&94.0&93.9&90.4&91.4\\
$\Delta \xi_{0,2,\alpha}$&94.6&94.5&90.6&92.7&95.1&94.6&90.8&91.2\\
$\Delta \xi_{0,3,\alpha}$&95.1&94.5&90.6&92.6&95.0&94.5&90.5&91.4\\
\hline
&\multicolumn{8}{c}{\footnotesize $(n_0,n_1,n_2,n_3)=(38,45,60,60),
~~(\sigma^2_{\gamma,0},\sigma^2_{\gamma,1},\sigma^2_{\gamma,2},\sigma^2_{\gamma,3})
=(1.44,1.44,1.00,1.00)$} \\
$\xi_{0,\alpha}$  &93.0&93.1&86.7&86.5&92.0&92.8&81.1&80.9\\
$\xi_{2,\alpha}$  &92.9&93.1&89.2&90.0&93.2&93.6&85.7&84.7\\
$\xi_{3,\alpha}$  &92.6&93.7&89.4&89.9&93.4&93.5&86.2&85.9\\
$\Delta \xi_{0,1,\alpha}$&94.4&94.4&88.2&88.3&94.2&94.3&82.2&82.0\\
$\Delta \xi_{0,2,\alpha}$&94.8&94.5&87.9&88.4&94.4&94.2&83.4&82.7\\
$\Delta \xi_{0,3,\alpha}$&95.0&94.6&88.6&88.0&94.3&94.3&83.3&83.1\\
\hline
&\multicolumn{8}{c}{\footnotesize $(n_0,n_1,n_2,n_3)=(38,45,60,60),
~~(\sigma^2_{\gamma,0},\sigma^2_{\gamma,1},\sigma^2_{\gamma,2},\sigma^2_{\gamma,3})
=(0.36,0.36,0.25,0.25)$}\\
$\xi_{0,\alpha}$  &93.2&93.5&89.1&90.9&93.2&93.3&88.9&89.8\\
$\xi_{2,\alpha}$  &93.4&93.8&92.0&92.9&93.7&94.1&91.7&92.2\\
$\xi_{3,\alpha}$  &93.4&93.8&91.9&93.0&93.9&94.0&91.7&91.9\\
$\Delta \xi_{0,1,\alpha}$&94.9&94.7&91.2&92.1&94.4&94.2&90.2&90.8\\
$\Delta \xi_{0,2,\alpha}$&94.8&94.7&91.1&92.2&94.7&94.9&90.2&91.0\\
$\Delta \xi_{0,3,\alpha}$&94.9&94.7&90.9&92.1&94.9&94.5&89.9&91.1\\
\hline
\end{tabular}

\end{center}
\end{table}

\begin{table}[h]
\centering\caption{Coverage probabilities (\%) of two-sided 95\% CIs under gamma random effects model.
Here bootstrap composite EL and Wald empirical quantile intervals:   CEL and EMP
\label{cover.pr.gam}}

\begin{center}
\tabcolsep 10pt
\renewcommand{\arraystretch}{1}

\begin{tabular}{c|rrrr rrrr}
\hline
&\multicolumn{4}{c }{$d=5$}&\multicolumn{4}{c}{$d=10$}\\
\hline
Method&\multicolumn{2}{c}{\mbox{CEL}}&\multicolumn{2}{c }{\mbox{EMP}}&\multicolumn{2}{c}{\mbox{CEL}}&\multicolumn{2}{c}{\mbox{EMP}}\\
$\alpha$ &0.05&0.10 &0.05&0.10  &0.05&0.10&0.05&0.10 \\
\hline
&\multicolumn{8}{c}{$(n_0,n_1,n_2,n_3)=(25,30,40,40),~~b=14$}\\
$\xi_{0,\alpha}$  &90.6&91.2&85.7&88.8&90.7&91.3&82.1&82.3\\
$\xi_{2,\alpha}$  &91.7&92.2&90.4&90.3&91.9&92.3&85.8&84.7\\
$\xi_{3,\alpha}$  &91.5&92.6&90.8&91.8&92.3&92.8&86.7&85.8\\
$\Delta \xi_{0,1,\alpha}$&93.9&93.9&87.3&89.1&94.2&94.3&83.5&82.8\\
$\Delta \xi_{0,2,\alpha}$&93.8&93.7&87.3&89.4&93.5&93.2&83.4&83.7\\
$\Delta \xi_{0,3,\alpha}$&93.9&93.7&87.7&89.5&93.4&93.3&83.7&84.0\\
\hline
&\multicolumn{8}{c}{$(n_0,n_1,n_2,n_3)=(25,30,40,40),~~b=63$}\\
$\xi_{0,\alpha}$  &92.0&92.3&90.3&93.6&92.1&92.3&90.5&92.4\\
$\xi_{2,\alpha}$  &92.7&93.1&93.4&94.6&93.2&93.1&92.5&92.8\\
$\xi_{3,\alpha}$  &92.9&93.8&93.9&94.9&93.5&93.8&93.1&93.2\\
$\Delta \xi_{0,1,\alpha}$&94.1&93.8&92.4&94.3&93.6&93.7&91.8&93.0\\
$\Delta \xi_{0,2,\alpha}$&94.1&94.2&92.4&94.8&94.3&94.2&91.0&92.9\\
$\Delta \xi_{0,3,\alpha}$&94.8&94.1&91.9&94.3&94.0&93.9&91.3&92.4\\
\hline
&\multicolumn{8}{c}{$(n_0,n_1,n_2,n_3)=(38,45,60,60),~~b=14$}\\
$\xi_{0,\alpha}$  &92.0&92.4&87.5&88.1&91.9&92.2&81.8&81.7\\
$\xi_{2,\alpha}$  &92.9&93.3&90.1&90.4&93.1&93.3&85.0&83.6\\
$\xi_{3,\alpha}$  &92.7&93.5&90.6&91.6&93.2&93.5&87.1&86.1\\
$\Delta \xi_{0,1,\alpha}$&94.0&94.1&88.4&89.4&94.0&94.1&83.2&81.7\\
$\Delta \xi_{0,2,\alpha}$&94.4&94.4&88.3&88.9&94.4&94.3&83.3&82.5\\
$\Delta \xi_{0,3,\alpha}$&94.3&93.9&88.6&89.4&94.0&94.3&83.2&82.2\\
\hline
&\multicolumn{8}{c}{$(n_0,n_1,n_2,n_3)=(38,45,60,60),~~b=63$}\\
$\xi_{0,\alpha}$  &93.3&93.5&91.9&93.3&92.8&92.8&90.3&92.2\\
$\xi_{2,\alpha}$  &93.3&93.7&93.0&94.3&94.0&94.1&92.5&92.9\\
$\xi_{3,\alpha}$  &93.6&94.1&93.8&95.1&94.0&94.1&93.0&93.7\\
$\Delta \xi_{0,1,\alpha}$&94.6&94.7&92.4&94.2&94.5&94.2&91.5&92.2\\
$\Delta \xi_{0,2,\alpha}$&95.1&95.1&92.7&93.7&94.3&94.4&91.0&92.3\\
$\Delta \xi_{0,3,\alpha}$&94.9&94.7&92.4&93.8&94.4&94.2&91.5&92.3\\
\hline
\end{tabular}

\end{center}
\end{table}

The Wald intervals have coverage probabilities much lower than the nominal 95\%.
This reveals the negative effect of ignoring the within-cluster correlations
(not due to the Wald method).
The bootstrap intervals (CEL) have coverage  probabilities much closer to 95\%.
The cluster-based bootstrapping method is clearly a good choice.

In all cases, the coverage probabilities of the bootstrap intervals are very close to 95\% for
the population quantile differences.
For individual population quantiles, the bootstrap method works well when the sample sizes
are large or the within-cluster correlation is low.
Otherwise, the coverage probability is as low as 90.6\% in
the most difficult case where the 5th population quantile is of interest, and
the sample size is low or the
random effect is high ($\sigma^2_{\gamma, 0} = 1.44$).
Improvement is desirable in these situations.

\subsection{Monitoring tests}

In a forestry project, clustered data on mechanical
strength are collected to {\blue monitor lumber quality over time.}
The specific monitoring target is the 5\% quantile, but the
problem is generic. Hence, we will also include the 50\% quantile,
namely, the median in the simulation.
In statistical terminology, we wish to test for the hypotheses
\[
H_{r,0}: \Delta\xi_{0,r,\alpha} \leq 0 \mbox{~versus~}
H^a_{r,0}:  \Delta\xi_{0,r,\alpha}> 0
\]
for some $r$ in $1, 2, \ldots, m$ and $\alpha$.
We use simulation to demonstrate that the
proposed method provides a highly effective monitoring tool.

We generate data from the same normal and  gamma random effects models.
Our settings make $\Delta\xi_{0, 1, \alpha} = 0$, but $\Delta\xi_{0,2,\alpha} > 0$
and  $\Delta\xi_{0,3,\alpha}> 0$ for any $\alpha$.
In other words, the data are generated from a model in which
$H_{1, 0}$ is true but $H_{2, 0}$ and $H_{3, 0}$ are false.

Existing methods are not readily applicable to monitoring tests.
CIs using the Wald method based on empirical quantiles
have undercoverage when the data are clustered, as shown in the last section.
Hence, this approach will not lead to a good monitoring test.
None of the made-to-measure methods included in \cite{verrill2015simulation}
work well.
They find that the most promising traditional
one-sided Wilcoxon test ($W$)
has inflated type I error when the data are clustered.
Interestingly, a rank-sum-test for clustered data
($W_c$) has been developed by  \cite{rosner2006wilcoxon}.
Because they are designed for different hypotheses, neither $W$ nor
$W_c$ serves the purpose of monitoring the 5\% or 50\% quantiles.
Nevertheless, we include both in the simulation.

Table \ref{power.normal} gives the simulation results for the models
defined in the previous section.
We use CEL$_{0.05}$ and CEL$_{0.5}$ to indicate our methods
monitoring the 5\% and 50\% quantiles.
First consider the rows labeled $H_{1, 0}$, which is
a true null hypothesis.
Rejection of $H_{1, 0}$ contributes to the type I error.
The standard Wilcoxon test ($W$) is clearly seen to have seriously
inflated type I errors.
The Wilcoxon test for clustered data ($W_c$) has precise
type I errors.
The proposed test has type I errors between
5.1\% and 6.2\%. The test for monitoring
the median (the 50\% quantile) is more precise
because there are more observations with sizes close to the median.

\begin{table}[h]
\centering\caption{Rejection rates (\%) for the $0.05$th and 0.5th quantiles under random effects models.
Here Wilcoxon tests: $W$, $W_c$; composite EL method: CEL$_{0.05}$, CEL$_{0.5}$; nominal level 5\%
\label{power.normal}}

\begin{center}
\tabcolsep 8pt
\renewcommand{\arraystretch}{1}

\begin{tabular}{c|cccc c cccc}
\hline
&\multicolumn{9}{c}{Normal random effects} \\
\hline
&\multicolumn{4}{c}{$d=5$}&&\multicolumn{4}{c}{$d=10$}\\
\hline
Method&CEL$_{0.05}$&CEL$_{0.5}$& W&$W_c$&  &CEL$_{0.05}$&CEL$_{0.5}$&W&$W_c$\\
\hline
&\multicolumn{9}{c}{\footnotesize $(n_0,n_1,n_2,n_3)=(25,30,40,40)$,
$(\sigma^2_{\gamma,0},\sigma^2_{\gamma,1},\sigma^2_{\gamma,2},\sigma^2_{\gamma,3})
=(1.44,1.44,1.00,1.00)$}\\
\hline
$H_{1,0}$&5.5  &5.5&12.1 & 5.2  &&6.2   &5.5&18.3     &5.6  \\
$H_{2,0}$&40.2&69.8& 83.5 & 68.8&&48.9 &78.1&93.0     &76.9 \\
$H_{3,0}$&83.8&99.3&99.8 & 99.1& &92.7&99.8&100.0   &99.8  \\
\hline
&\multicolumn{9}{c}{\footnotesize $(n_0,n_1,n_2,n_3)=(25,30,40,40)$, $
(\sigma^2_{\gamma,0},\sigma^2_{\gamma,1},\sigma^2_{\gamma,2},\sigma^2_{\gamma,3})
=(0.36,0.36,0.25,0.25)$}\\
\hline
$H_{1,0}$&5.2   &5.6 &7.7     &5.3  &&5.9&5.4  &10.4   & 5.2    \\
$H_{2,0}$&62.6 &90.4&92.7   &89.5 &&82.1&97.8&99.2   &97.6  \\
$H_{3,0}$&97.3 &100.0&100.0 &100 &&99.9&100.0&100.0 &100.0    \\
\hline
&\multicolumn{9}{c}{\footnotesize $(n_0,n_1,n_2,n_3)=(38,45,60,60),~~
(\sigma^2_{\gamma,0},\sigma^2_{\gamma,1},\sigma^2_{\gamma,2},\sigma^2_{\gamma,3})
=(1.44,1.44,1.00,1.00)$}\\
\hline
$H_{1,0}$&5.7&5.5&13.1     &4.9 &&6.0 &5.3 &19.1   & 5.0     \\
$H_{2,0}$&49.1&85.4&92.7   &83.7&&60.3&90.9&98.0  &91.2  \\
$H_{3,0}$&93.7&100.0&100.0 &100&&97.9&100.0&100.0 &100.0   \\
\hline
&\multicolumn{9}{c}{\footnotesize $(n_0,n_1,n_2,n_3)=(38,45,60,60)$, $
(\sigma^2_{\gamma,0},\sigma^2_{\gamma,1},\sigma^2_{\gamma,2},\sigma^2_{\gamma,3})
=(0.36,0.36,0.25,0.25)$}\\
$H_{1,0}$&5.1  &5.2& 7.8     &5.2 &&5.7&5.3&10.4      &5.6  \\
$H_{2,0}$&76.4&97.9&98.5   &97.2&&92.7&99.8&100.0  &99.8  \\
$H_{3,0}$&99.7&100.0&100.0 &100&&100.0&100.0&100.0&100.0   \\
\hline
&\multicolumn{9}{c}{Gamma random effects}\\
\hline
%&\multicolumn{4}{c}{$d=5$}&&\multicolumn{4}{c}{$d=10$}\\
%\hline
%Method&CEL&CEL$_m$& W&$W_c$&  &CEL&CEL$_m$&W&$W_c$\\
&\multicolumn{7}{c}{\footnotesize $(n_0,n_1,n_2,n_3)=(25,30,40,40),~~b=14$}\\
$H_{0,1}$&6.0   &5.5 &14.1  &  5.6 &&5.9   &5.6  &21.1  &5.3   \\
$H_{0,2}$&74.2 &89.2&96.0  & 88.4 &&84.5&92.8  &99.0  &92.2   \\
$H_{0,3}$&99.8 &100.0&100.0& 100.0 &&100.0&100.0&100.0&100.0    \\
\hline
&\multicolumn{7}{c}{\footnotesize $(n_0,n_1,n_2,n_3)=(25,30,40,40),~~b=63$}\\
$H_{0,1}$&5.9    &5.3     & 8.2    &4.9  &&6.0&5.2    &11.7   &5.2     \\
$H_{0,2}$&85.9  &98.5   & 98.7  &97.6&&97.4&99.7  &100.0&  99.7 \\
$H_{0,3}$&100.0&100.0 & 100.0&100.0 &&100.0&100.0&100.0&  100.0     \\
\hline
&\multicolumn{7}{c}{\footnotesize $(n_0,n_1,n_2,n_3)=(38,45,60,60),~~b=14$}\\
$H_{0,1}$&5.9    &4.7    &14.1   &5.4    &&5.7  &6.2  & 21.5  & 5.1   \\
$H_{0,2}$&87.0  &97.0  &99.2   &96.8  &&93.4&98.2  &99.9   &  98.4   \\
$H_{0,3}$&100.0&100.0&100.0 &100.0   &&100.0&100.0&100.0 &  100.0   \\
\hline
&\multicolumn{7}{c}{\footnotesize $(n_0,n_1,n_2,n_3)=(38,45,60,60),~~b=63$}\\
$H_{0,1}$&5.7     &5.0   &8.8    &5.3  &&5.8 &5.1   &11.5  & 4.8   \\
$H_{0,2}$&95.9   &99.8 &99.9  &99.7 &&99.8 &100.0 &100.0& 100.0    \\
$H_{0,3}$&100.0 &100.0&100.0&100.0 &&100.0&100.0&100.0& 100.0    \\
\hline
\end{tabular}

\end{center}
\end{table}

Next, consider the rows labeled $H_{2,0}$ and $H_{3,0}$,
which are false hypotheses. The rejection rates of the
proposed CEL and $W_c$ are markedly larger than the type I errors,
and they increase with the sample sizes. Hence, the simulation
fully supports the effectiveness of the proposed method.
It also appears that $W_c$ is a respectable choice.

The simulation results show that the power for monitoring
the 5\% quantile of the proposed method is much lower than
that for the median. A comparison with $W_c$
is difficult because it does not monitor changes in the quantiles.
As \cite{kruskal1952nonparametric} points out, the Wilcoxon test was introduced
to detect/monitor location shifts, but it actually monitors
\be
\label{Wilcoxon-H}
H'_0: \p(X_0<X_1)=0.5 \mbox{ versus } H'_a: \p(X_0<X_1)\neq 0.5
\ee
where $X_0$ and $X_1$ are independent random variables
representing two populations.

The success of the Wilcoxon test comes from the fact that
a location shift often coincides with a shift in $\p(X_0<X_1)$.
For the same reason, $W_c$ can detect changes in the quantile
due to veracities of $H_0'$ and $H_0$.
When a shift in the quantile does not coincide with
a change in $\p(X_0<X_1)$, $W_c$ becomes useless.
We give two examples as follows.
\begin{enumerate}
\item[M1]
Gamma random effects model with 2 populations and parameter values
$(a_0, a_1)=(8,16)$, $b=63$, and $(\beta_0, \beta_1)=(1.05, 2.511)$.
Sample sizes are $40, 40$ with $d=10$.

Note that $\xi_{0, 0.05} - \xi_{1, 0.05} < 0$,
$\xi_{0, 0.1} - \xi_{1, 0.1} = 0$, but $\p(X_0<X_1)<0.5$.

\item[M2]
Normal random effects model with 2 populations and parameter values
$(\mu_0,\mu_1) =(15.5, 15.5)$,
$(\sigma^2_{\gamma,0}, \sigma^2_{\epsilon,0})=(0.1, 0.9)$,
and $(\sigma^2_{\gamma,1}, \sigma^2_{\epsilon,1})=(0.2, 1.8)$.
Sample sizes are also $40, 40$ with $d=10$.

Note that
$\xi_{0,0.05} - \xi_{1,0.05} > 0$, $\xi_{0,0.1}-\xi_{1,0.1}>0$, and $\p(X_0 < X_1)=0.5$.
\end{enumerate}

Under M1, both
\(
H_0^{(1)}: \Delta \xi_{0,1,0.05} \leq  0
\)
and
\(
H_0^{(2)}: \Delta \xi_{0,1,0.10} \leq  0
\)
hold. Under M2, both are violated. Hence, for monitoring
the $5\%$ or $10\%$ quantiles,  a good test
should have rejection rates that are below 5\% under M1 and
high under M2.
We use 10,000 repetitions in the simulation
so that the simulation error is below $0.5\%$;
Table \ref{specialExample} gives the simulated rejection rates.
Clearly, the proposed monitoring test serves the monitoring purpose as promised, but
the Wilcoxon tests, whether or not they are designed for clustered data, do not.

\begin{table}[h]
\centering\caption{Rejection rates (\%)  under two special models
\label{specialExample}}

\begin{center}
\tabcolsep 8pt
\renewcommand{\arraystretch}{1}

\begin{tabular}{c|ccccccc}
\hline
& \multicolumn{3}{c}{$H_0^{(1)}$} &&  \multicolumn{3}{c}{$H_0^{(2)}$}\\
      & CEL$_{0.05}$ & $W$ & $W_c$ &  & CEL$_{0.10}$ & $W$ &  $W_c$\\ \hline
M1 &  3.7  & 99.0 & 99.0        &  & 4.84 & 99.0 & 99.0 \\
M2 & 99.5 & 11.3 & 5.3    &  & 97.4 & 11.3 & 5.3 \\
\hline
\end{tabular}

\end{center}
\end{table}

\subsection{Overfitted or misspecified density ratio model}

In applications, practitioners do not have the luxury of knowing the most
suitable basis function $\bq(\cdot)$. To control the risk of misspecification of the
DRM, one may intentionally choose an extensive basis
vector function $\bq(\cdot)$.

We use  additional simulation experiments to show that this strategy works.
We consider two situations. The DRM is still correct  but overfitted in the first situation,
and is  misspecified in the second situation.
For the first situation, we generate data using the model settings in the last section.
We use $\bq(y) = (1, y, y^2, \log y, \log^2 y)^\tau$ for the data from both models.
To save space, we report in Table \ref{overfit}
only the type I errors and powers of the proposed test.
For comparison, the corresponding figures obtained using
a nonredundant $\bq(y)$ are also included in the table.
The variances under the normal random effects model are
$(\sigma^2_{\gamma,0},\sigma^2_{\gamma,1},\sigma^2_{\gamma,2},
\sigma^2_{\gamma,3})=(0.36,0.36,0.25,0.25)$, and the
parameter under the gamma random effects model is
$b = 63$. The results for the other parameter settings are similar.

It is clear that using the overfitted basis function
$\bq(y) = (1, y, y^2, \log y, \log^2y)^\tau$ leads to
better controlled type I errors in all cases with negligible
loss of power. The viability of the proposed method
is therefore well supported.

\begin{table}[h]
\centering\caption{Rejection rates (\%) of CEL monitoring test under overfitted DRMs.
Here nominal level: 5\%;  cluster size $d=10$; $\alpha$: level of quantile
\label{overfit}}

\begin{center}
\tabcolsep 8pt
\renewcommand{\arraystretch}{1}

\begin{tabular}{ccccccccc}
\hline
&\multicolumn{4}{c}{Normal random effects}
&\multicolumn{4}{c}{Gamma random effects}\\
$\bq(y)$ &\multicolumn{2}{c}{$(1,y,y^2)^\tau$}
&\multicolumn{2}{c}{$(1, y, y^2, \log y, \log^2y)^\tau$}
&\multicolumn{2}{c}{$(1, \log y, y)^\tau$}
&\multicolumn{2}{c}{$(1, y, y^2,\log y, \log^2y)^\tau$} \\
\hline
$\alpha$ &$0.05$&$0.1$&$0.05$&$0.1$&$0.05$&$0.1$&$0.05$&$0.1$\\
\hline
&\multicolumn{8}{c}{$(n_0,n_1,n_2,n_3)=(25,30,40,40)$}\\
$H_{1,0}$&5.9&5.9&5.3&5.8 &6.0&6.0&5.7&5.9\\
$H_{2,0}$&82.1&89.0&78.8&88.1&97.4&98.8&95.8&98.6\\
$H_{3,0}$&99.9&100.0&99.7&100.0&100.0&100.0&100.0&100.0\\
&\multicolumn{8}{c}{$(n_0,n_1,n_2,n_3)=(38,45,60,60)$}\\
$H_{1,0}$&5.7&5.5&5.4&5.4&5.8&5.8&5.8&5.6\\
$H_{2,0}$&92.7&96.9&89.8&96.5&99.8&100.0&99.4&99.9\\
$H_{3,0}$&100.0&100.0&100.0&100.0&100.0&100.0&100.0&100.0\\
\hline
\end{tabular}

\end{center}
\end{table}

For the second situation, we generate data from the following two models.

\noindent{\it \bf Model 3: Weibull random effects model.}
The data is generated as follows:
\begin{equation}
\label{weibull.mixed}
y_{k,j,l}=\gamma_{k,j}\epsilon_{k,j,l},
\end{equation}
where $\epsilon_{k,j,1},\ldots,\epsilon_{k,j,d}$ are $d$ {\iid} random variables
from beta distribution with shape parameters $a_k$ and $b_k$,
$\gamma_{k,1},\ldots,\gamma_{k,n_k}$ are $n_k$ {\iid} random variables from
weibull distribution with shape parameter $c_{k}$ and scale parameter $d_k$.
Here a weibull distribution with shape parameter $c$ and scale parameter $d$
has the density function
$$
\frac{c}{d}\left(\frac{x}{d}\right)^{c-1}\exp\left\{-\left(\frac{x}{d}\right)^{c} \right\} \ind(0 < x).
$$

We choose $m+1 = 4$ populations with the parameter values
\[
a_0= a_1 = 8, ~a_2 = 7, ~a_3 = 6; ~~
b_0=b_1=b_2=b_3 =63
\]
and
\[
c_0=c_1=c_2=c_3 =10;~~
d_0=d_1=74,~d_2=70,~d_3=66.
\]

\noindent{\it \bf Model 4: Mixed gamma and weibull  random effects model.}
In this model, we still consider $m+1 = 4$ populations.
The first two populations are generated from gamma random effects model
with
$$
a_0= a_1 = 8;~~b_0=b_1=63;~~\beta_0=\beta_1=1
$$
and
the last two populations are generated from weibull random effects model
with
\[
a_2 = 7, ~a_3 = 6; ~~
b_2=b_3 =63;~~c_2=c_3 =10;~~d_2=70,~d_3=66.
\]

For both models, we set $d=10$, and consider  $(n_0,n_1,n_2,n_3)=(25,30,40,40)$ and $(n_0,n_1,n_2,n_3)=(38,45,60,60)$.
Neither Model 3 nor Model 4  satisfies the DRM.
However, we still fit the data from both models by the DRM with  $\bq(y) = (1, y, y^2, \log y, \log^2 y)^\tau$.

We use 10,000 repetitions to obtain
the {\amse}s
of the CEL quantiles and the EMP quantiles or their differences across the
four populations, the coverage probabilities of the two-sided 95\% cluster-based bootstrap CIs based on the CEL quantiles and
two-sided 95\% Wald-type  CIs based on the EMP quantiles,
and
the rejection rates of the proposed monitoring test for
testing
\[
H_{r,0}: \Delta\xi_{0,r,\alpha} \leq 0 \mbox{~versus~}
H^a_{r,0}:  \Delta\xi_{0,r,\alpha}> 0
\]
for $r=1, 2, 3$ and $\alpha=0.05,0.10$.
The simulation results
are presented in Tables \ref{compare.mse3}--\ref{power.mis}.

%\centerline{Put Tables \ref{compare.mse3}--\ref{power.mis} about here}

\begin{table}[h]
\centering\caption{\amse\ ($\times$100)  of the composite EL and empirical quantiles (CEL and EMP) under misspecified DRMs.
Here cluster size $d=10$; $\Delta \xi_{0,k,\alpha} = \xi_{0,\alpha} - \xi_{k,\alpha}$
\label{compare.mse3}}

\begin{center}
\tabcolsep 8pt
\renewcommand{\arraystretch}{1}

\begin{tabular}{crrrrrrrr}
\hline
&\multicolumn{4}{c }{Model 3}&\multicolumn{4}{c}{Model 4}\\
\hline
Method&\multicolumn{2}{c}{\mbox{CEL}}&\multicolumn{2}{c }{\mbox{EMP}}&\multicolumn{2}{c}{\mbox{CEL}}&\multicolumn{2}{c}{\mbox{EMP}}\\
$\alpha$ &0.05&0.10 &0.05&0.10  &0.05&0.10&0.05&0.10 \\
\hline
&\multicolumn{8}{c}{$(n_0,n_1,n_2,n_3)=(25,30,40,40)$}\\
$\xi_{0,\alpha}$             &5.79&5.53&7.42&6.68&4.92&4.67&6.57&5.70\\
$\xi_{2,\alpha}$              &2.66&2.52&3.27&2.92&2.74&2.58&3.40&3.03\\
$\xi_{3,\alpha}$              &1.98&1.82&2.17&2.00&2.03&1.89&2.22&2.05\\
$\Delta \xi_{0,1,\alpha}$&10.45&10.00&14.08&12.37&8.55&8.29&11.96&10.56\\
$\Delta \xi_{0,2,\alpha}$&8.33&8.09&10.73&9.67&7.66&7.31&10.02&8.82\\
$\Delta \xi_{0,3,\alpha}$&7.91&7.48&9.77&8.72&6.96&6.62&8.78&7.76\\
\hline
&\multicolumn{8}{c}{$(n_0,n_1,n_2,n_3)=(38,45,60,60)$}\\
$\xi_{0,\alpha}$               &3.94&3.72&5.18&4.49&3.17&3.05&4.23&3.84\\
$\xi_{2,\alpha}$              &1.78&1.72&2.21&2.00&1.79&1.71&2.16&2.01\\
$\xi_{3,\alpha}$              &1.33&1.24&1.47&1.37&1.33&1.27&1.47&1.38\\
$\Delta \xi_{0,1,\alpha}$&6.97&6.63&9.24&8.09&5.59&5.39&7.70&7.00\\
$\Delta \xi_{0,2,\alpha}$&5.69&5.45&7.47&6.56&4.93&4.80&6.49&5.92\\
$\Delta \xi_{0,3,\alpha}$&5.33&5.03&6.73&5.88&4.52&4.36&5.69&5.25\\
\hline
\end{tabular}

\end{center}
\end{table}

\begin{table}[h]
\centering\caption{Coverage probabilities (\%) of two-sided 95\% CIs under misspecified DRMs.
Here bootstrap composite EL and Wald empirical quantile intervals: CEL and EMP
\label{cover.mis}}

\begin{center}
\tabcolsep 8pt
\renewcommand{\arraystretch}{1}

\begin{tabular}{crrrrrrrr}
\hline
&\multicolumn{4}{c }{Model 3}&\multicolumn{4}{c}{Model 4}\\
\hline
Method&\multicolumn{2}{c}{\mbox{CEL}}&\multicolumn{2}{c }{\mbox{EMP}}&\multicolumn{2}{c}{\mbox{CEL}}&\multicolumn{2}{c}{\mbox{EMP}}\\
\hline
&\multicolumn{8}{c}{$(n_0,n_1,n_2,n_3)=(25,30,40,40)$}\\
$\xi_{0,\alpha}$             &90.0&91.5&88.6&90.2&90.9&92.1&90.1&92.4\\
$\xi_{2,\alpha}$             &92.2&93.0&90.7&91.4&92.0&93.2&90.0&90.6\\
$\xi_{3,\alpha}$             &93.2&93.2&91.8&92.3&92.9&93.1&91.6&92.4\\
$\Delta \xi_{0,1,\alpha}$&93.2&93.6&89.1&90.3&93.2&93.6&91.5&92.6\\
$\Delta \xi_{0,2,\alpha}$&93.4&93.6&89.6&90.7&93.4&93.8&90.4&91.8\\
$\Delta \xi_{0,3,\alpha}$&92.8&93.0&89.2&90.5&93.3&93.6&90.7&92.2\\
\hline
&\multicolumn{8}{c}{$(n_0,n_1,n_2,n_3)=(38,45,60,60)$}\\
$\xi_{0,\alpha}$             &91.2&92.4&87.0&89.3&92.1&93.6&90.3&91.2\\
$\xi_{2,\alpha}$             &93.1&93.6&90.8&91.2&92.5&93.5&91.0&91.0\\
$\xi_{3,\alpha}$             &93.5&94.2&91.8&92.0&93.8&94.1&91.6&92.0\\
$\Delta \xi_{0,1,\alpha}$&93.8&94.0&88.9&90.0&93.9&94.1&91.8&91.9\\
$\Delta \xi_{0,2,\alpha}$&93.6&93.7&88.3&89.6&94.0&94.5&91.0&91.2\\
$\Delta \xi_{0,3,\alpha}$&93.3&93.6&88.5&89.9&94.0&94.4&91.2&91.8\\
\hline
\end{tabular}

\end{center}
\end{table}

\begin{table}[h]
\centering\caption{Rejection rates (\%) of CEL monitoring test under misspecified DRMs.
Here nominal level: 5\%;  cluster size $d=10$; $\alpha$: level of quantile
\label{power.mis}}

\begin{center}
\tabcolsep 8pt
\renewcommand{\arraystretch}{1}

\begin{tabular}{lrrrr}
\hline
&\multicolumn{2}{c}{Model 3}
&\multicolumn{2}{c}{Model 4}\\
\hline
$\alpha$ &$0.05$&$0.1$&$0.05$&$0.1$\\
\hline
&\multicolumn{4}{c}{$(n_0,n_1,n_2,n_3)=(25,30,40,40)$}\\
$H_{1,0}$&6.2&6.0&6.0&6.0\\
$H_{2,0}$&90.5&94.6&96.3&98.5\\
$H_{3,0}$&99.9&100.0&100.0&100.0\\
\hline
&\multicolumn{4}{c}{$(n_0,n_1,n_2,n_3)=(38,45,60,60)$}\\
$H_{1,0}$&5.8&5.8&5.8&5.7\\
$H_{2,0}$&96.9&98.8&99.4&99.9\\
$H_{3,0}$&100.0&100.0&100.0&100.0\\
\hline
\end{tabular}

\end{center}
\end{table}

From Table \ref{compare.mse3}, we see that the CEL quantiles are uniformly more efficient than the EMP quantiles.
Sometimes, the efficiency gain can be as high as 25\%; see for example the estimation of $\Delta\xi_{0,1,0.05}$
under both Models 3 and 4.
We have also calculated the bias for the CEL quantiles.
The maximum value of the absolute bias in all cases are less than 0.03,
which are negligible.
To save the space, the biases of CEL quantiles and EMP quantiles are not reported.
From Table \ref{cover.mis}, we see that  the cluster-based bootstrap CIs
still have quite close to the nominal coverages.
Further, they have more accurate coverage than the Wald-type  CIs based on the EMP quantiles.
Table \ref{power.mis} indicates that the proposed monitoring tests have quite accurate type I errors although the DRM is misspecified.

\section{Illustrative application}

In this section, we apply the proposed bootstrap CEL
monitoring test to a real data set.
The data set contains two samples from two populations that
will be referred to as In-Grade and 2011/2012.
The In-Grade sample consists of 398 modulus of rupture (MOR) measurements.
They are collected from lumber grades as commercially produced.
The 2011/2012 sample consists of 408 MOR measurements.

For the In-Grade samples, MOR measurements are obtained from 27 mills:
14 mills sampled 10 pieces from a single lot;
2 mills sampled 9 pieces from one lot and 10 pieces from another;
and 11 mills sampled 10 pieces from each of two lots.
For the 2011/2012 samples, MOR measurements are obtained from 41 mills:
39 mills sampled 10 pieces
and  2 mills sampled 9 pieces from a single lot.
Apparently, the original plan was to have 10 pieces from each lot in the sample.
We use this data set
to conduct a monitoring test for the 5\% and 10\% quantiles of the MOR.

We first confirm the nonignorable random effects through a standard analysis
of variance (ANOVA) procedure \citep[pp 71--72]{wu2011experiments}
under the random effects model (\ref{random.effect}).
The null and alternative hypotheses are
$$
H_0:\sigma^2_{\gamma}=0 \mbox{ versus } H_a: \sigma^2_{\gamma} > 0.
$$
We used the R-function {\tt\blue aov} for this purpose, and the results are given
in Table \ref{aov}. The presence of random effects in both populations
is highly significant. The variance of the random effect is estimated as
$\hat \sigma^2_\gamma = 0.3 $ for both populations,
and the error variances are estimated as {\blue
$\hat \sigma^2_\epsilon = 4.3$ and 3.0, respectively.}
Their relative sizes are matched in the models used in the simulation.

\begin{table}[h]
\centering\caption{ANOVA table based on In-Grade sample and 2011/2012 sample
\label{aov} }

\begin{center}
\tabcolsep 8pt
\renewcommand{\arraystretch}{1}

\begin{tabular}{c|ccccc}
\hline
In-Grade  sample & Df & Sum Sq & Mean Sq &F-value&  P-value \\
                  \hline
Factor (lot)  & 39  &290.8  & 7.455  & 1.733 &0.006\\
Residuals  & 358 &1539.8   &4.301   \\
\hline \hline
2011/2012  sample & Df & Sum Sq & Mean Sq &F-value&  P-value \\
Factor (lot) & 40  &238.8   &5.970   &1.998 &0.001\\
Residuals  &  367 &1096.5  & 2.988   \\
\hline
\end{tabular}

\end{center}
\end{table}

The normality assumption in ANOVA is not crucial for detecting the
random effects. An analysis of the log-transformed data gives us
equally strong evidence for the existence of the nonignorable
random effects.

We recommend that the basis function vector $\bq(y) = (1, \log y)^\tau$
be used in the DRM for the bootstrap monitoring test.
Figure \ref{cdf_estimate} shows the corresponding fitted population distribution functions
$\hat G_0(y)$ and $\hat G_1(y)$ under the DRM together with
the empirical distribution functions $\tilde G_0(y)$ and $\tilde G_1(y)$.
Clearly, the DRM with this $\bq(y)$
fits these two populations well.
Other choices such as $(1, \log y, y)^\tau$
and $(1,\log y, \log^2y)^\tau$ are also adequate. We
will selectively present some of these results; the conclusions are nearly identical
in terms of quantile estimation and monitoring test.

\begin{figure}[h]
  \centering
\includegraphics [angle=-90, scale=0.4]{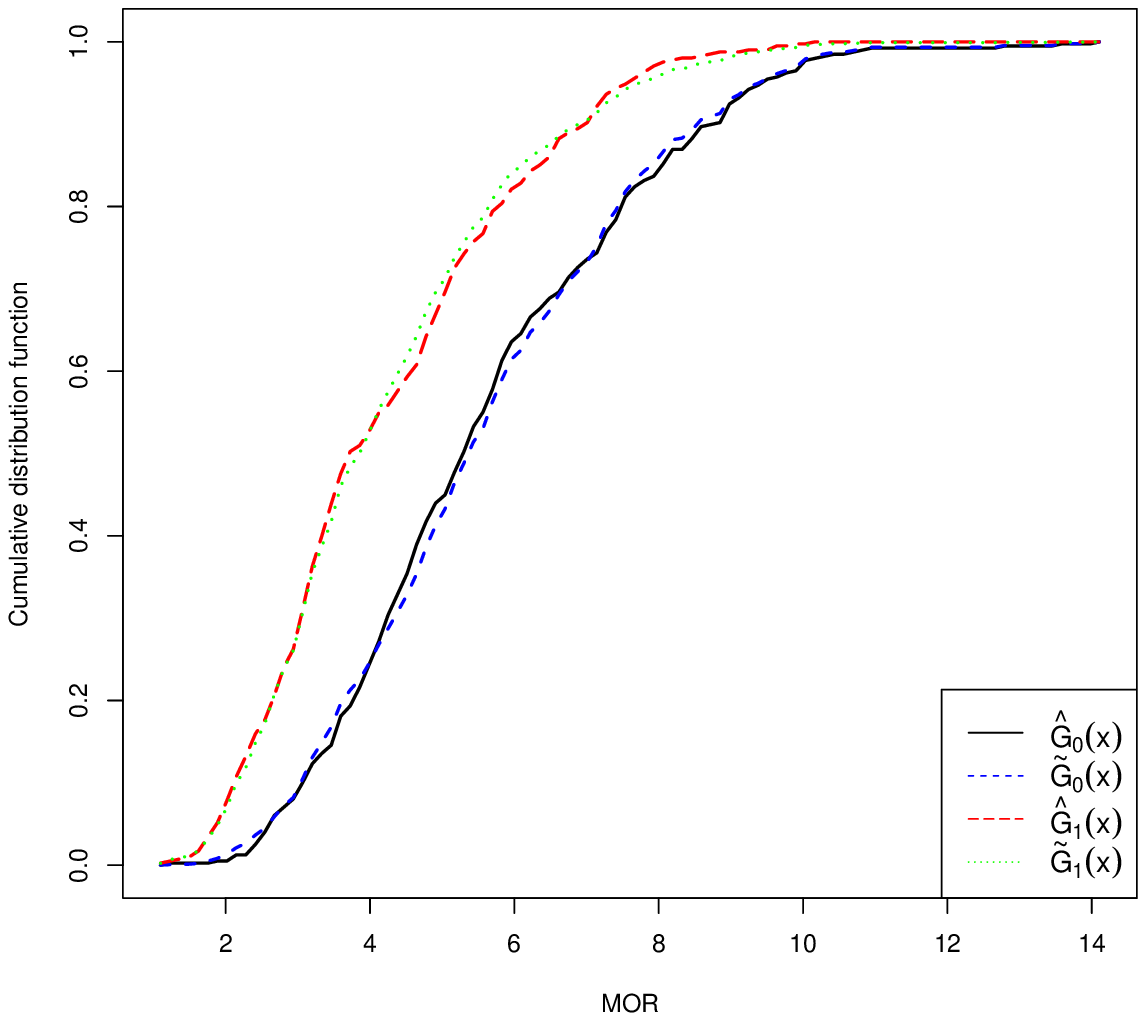}
  \caption{Fitted population distributions.
Here $\hat G_0(y)$ and $\hat G_1(y)$: fitted {\cdf} by DRM-CEL;
$\tilde G_0(y)$ and $\tilde G_1(y)$: empirical {\cdf}. }
\label{cdf_estimate}
\end{figure}

Most cluster sizes are $d=10$ with few exceptions in the actual data.
The proposed monitoring test can be carried out without difficulty;
Table \ref{oneside} includes all the information needed for the test.
Clearly, the data analysis leads to solid evidence against both
$
H_0:\xi_{0, 0.05}-\xi_{1, 0.05}\leq0
$
and
$
H_0: \xi_{0, 0.1}-\xi_{1, 0.1}\leq0
$
in favor of the one-sided alternatives $
H_a:\xi_{0, 0.05}-\xi_{1, 0.05} > 0
$
and
$
H_a: \xi_{0, 0.1}-\xi_{1, 0.1} > 0
$.
We confidently conclude that the 2011/2012 population
has lower quality index values than the In-Grade
population. Based on the theory developed in this paper, the risk of a false
alarm based on this analysis is low.

%\centerline{Put Table \ref{oneside} about here}

\begin{table}[h]
\centering\caption{Composite EL estimates and bootstrap confidence intervals for
$\Delta\xi_{0,1,\alpha}=\xi_{0,\alpha}-\xi_{1,\alpha}$
\label{oneside}}

\begin{center}
\tabcolsep 8pt
\renewcommand{\arraystretch}{1}

\begin{tabular}{c|cccccc}
\hline
$\bq(y)$&\multicolumn{2}{c}{Point Estimate}&\multicolumn{2}{c}{95\%
one-sided CI}&\multicolumn{2}{c}{99\% one-sided CI}\\
 &$\Delta\xi_{0,1,0.05}$&$\Delta\xi_{0,1,0.10}$& $\Delta\xi_{0,1,0.05}$&$\Delta\xi_{0,1,0.10}$& $\Delta\xi_{0,1,0.05}$&$\Delta\xi_{0,1,0.10}$\\
\hline
$(1,\log y)^\tau$& 0.677&0.903&$[0.508,\infty)$&$[0.650,\infty)$&$[0.438,\infty)$&$[0.563,\infty)$\\
$(1,y, \log y)^\tau$&0.695&0.916&$[0.497,\infty)$&$[0.642,\infty)$&$[0.416,\infty)$&$[0.535,\infty)$\\
$(1,\log y, \log^2y)^\tau$&0.734&0.922&$[0.515,\infty)$&$[0.659,\infty)$&$[0.429,\infty)$&$[0.552,\infty)$\\
\hline
\end{tabular}

\end{center}
\end{table}

\section{Summary and discussion}

 We have presented a DRM-based CEL approach
to analyze multiple samples containing clustered data.
The CEL is effective {\blue and} the cluster-based bootstrap
CIs have satisfactorily precise coverage probabilities.
Its derived monitoring test controls the type I error rates tightly
with good power.
We have shown these points through simulation studies
and a data example.
Further improvements in the precision of the
coverage probability and type I error rates
are possible. In the future, we aim to refine the current results along the
lines of \cite{loh1991bootstrap} and \cite{ho2005iterated}.

\section*{Supplementary Materials}

The online supplementary material includes the proofs of Theorems \ref{varcov2}--\ref{varxi2}.

%\section*{Acknowledgement}
%We  thank the editor, the AE, and  two anonymous referees for their constructive suggestions that significantly improved the paper.

\section*{Disclosure statement}

 No potential conflict of interest was reported by the authors.

\section*{Funding}

The authors gratefully acknowledge funding from the ``a thousand talents'' program
through Yunnan University and from NSERC Grants RGPIN-2014-03743 and  RGPIN-2015-06592,
the National Natural Science Foundation of China (Numbers  11771144 and 11371142),
and  the 111 Project (B14019),
a Collaborative Research and Development Grant from NSERC and FPInnovations.
We are indebted to Drs.~Steve Verrill, David Kretschmann, and James Evans at the
USDA Forest Products Lab for making their report available
as well as for providing the dataset on which their analyses and now ours are based.
We are also indebted to the Forest Products Stochastic Modelling Group
centered at the University of British Columbia (UBC): members of this group from
FPInnovations in Vancouver,  Simon Fraser University, and UBC provided stimulating discussions of the long-term
monitoring program to which this paper contributes.  Liu is the corresponding author.

%\section*{References}
%\bibliographystyle{chicago}      % Chicago style, author-year citations
%\bibliography{Jiahua2017}

\end{document}